# Large-scale database analysis of anomalous thermal conductivity of quasicrystals and its application to thermal diodes

**Takashi Kurono, Jinjia Zhang, Yasushi Kamimura, and Keiichi Edagawa[a)]**

**AFFILIATIONS**

Institute of Industrial Science, The University of Tokyo, Tokyo 153-8505, Japan

[a)] **Author to whom correspondence should be addressed:** edagawa@iis.u-tokyo.ac.jp

## ABSTRACT

One long-standing and crucial issues in the study of quasicrystals has been to identify the physical properties characteristic of quasicrystals. The large positive temperature coefficient of thermal conductivity at temperatures above room temperature, which has been observed in several quasicrystals, is one such characteristic property. Here, we show that this is indeed a very distinct property of quasicrystals through analysis using a large physical property database “Starrydata”. In fact, several quasicrystals ranked nearly first among more than 10,000 samples of various materials (metallic alloys, semiconductors, ceramics, etc.) in terms of the magnitude of the positive temperature coefficient of thermal conductivity. This unique property makes quasicrystals ideal for use in composite thermal diodes. We searched the database for the most suitable materials that can be combined with quasicrystals to create high-performance composite thermal diodes. Analytical calculations using a simple one-dimensional model showed that by selecting the optimal material, a thermal rectification ratio of 3.2 can be obtained. Heat transfer simulations based on the finite element method confirmed that this can be achieved under realistic conditions. This is the highest value of the thermal rectification ratio reported to date for this type of thermal diode.

## I. INTRODUCTION

Quasicrystals (QCs) are solid phases with a long-range structural order that is essentially different from that of conventional crystals.[1-3] QCs have so far been found to form in about 100 alloy systems in a thermodynamically stable state.[4] Since their discovery in 1984,[5] many studies have been conducted to reveal the physical properties characteristic of QC alloys. One such characteristic property is the large positive temperature coefficient of thermal conductivity above room temperature, which has been reported for several QC alloys.[6-10] The origin of this unusual property has been discussed in relation to the pseudogap in the electronic density of states,[11,12] often seen for QC alloys. Considering this property, Takeuchi et al. attempted to realize a high-performance thermal diode using QC alloys.[6,12,13]

Waste heat is the main source of global energy loss in modern society, and its reduction and/or recovery is crucial for maintaining sustainability. In this regard, technologies for controlling heat transport have attracted considerable attention, and the thermal diode is one of the key devices that can control the direction of heat transfer.[14,15] Since Starr first observed a thermal rectification effect in copper oxide in 1936,[16] various types of thermal diodes have been proposed, discussed, and experimentally demonstrated.[14,15] Among these, a composite thermal diode consisting of two bulk materials with positive and negative temperature coefficients of thermal conductivity[17,18] is advantageous for practical applications in many respects.[12]

The structure of a composite thermal diode is schematically illustrated in Fig. 1,[17,18] where materials A and B with positive and negative thermal conductivity temperature coefficients, respectively, are connected in series. Both ends of the diode are connected to thermal reservoirs having temperatures of $T_{\mathrm{H}}$ and $T_{\mathrm{L}}$ ($T_{\mathrm{H}} > T_{\mathrm{L}}$), respectively. When material A (B) is on the side of $T_{\mathrm{H}}$ ($T_{\mathrm{L}}$) (forward configuration), the thermal conductivities of both materials are relatively high, resulting in a relatively large heat flux $|J_{\mathrm{for}}|$ (Fig. 1 (a)). When material A (B) is on the side of $T_{\mathrm{L}}$ ($T_{\mathrm{H}}$) (reverse configuration), both materials have relatively low thermal conductivities, leading to a relatively small heat flux $|J_{\mathrm{rev}}|$ (Fig. 1 (b)). The performance of this diode can be evaluated using the thermal rectification ratio (TRR), which is defined as TRR=$|J_{\mathrm{for}}|/|J_{\mathrm{rev}}|$. To achieve a high TRR, two materials with large positive and negative temperature coefficients of thermal conductivity are required. Takeuchi et al. achieved a TRR=2.2 with $T_{\mathrm{H}}$ =900 K and $T_{\mathrm{L}}$ =300 K for a composite thermal diode with a QC alloy as material A in Fig. 1 by appropriately choosing a crystalline phase as material B.[12] They also investigated the combinations of materials not including QC alloys and achieved a TRR=2.7 with $T_{\mathrm{H}} = 413$ K and $T_{\mathrm{L}}$ =300 K for materials exhibiting a sharp change in thermal conductivity by phase transition.[19] To the best of our knowledge, this is the highest TRR ever reported for a composite thermal diode.

Katsura et al. have been constructing a large physical property database called "Starrydata",[20] which contained 184,838 *x*-*y* plotted curves collected from published papers as of April 4, 2024. This includes curves for various physical properties of solids, such as the temperature dependence of the electrical resistivity/conductivity, thermal conductivity, Seebeck coefficient and magnetic susceptibility, and the magnetic field dependence of the magnetization, etc. In this study, we analyzed 10,112 temperature dependence curves of thermal conductivity in Starrydata to determine the degree of quasicrystal specificity with respect to the large positive temperature coefficient of thermal conductivity above room temperature. The analyses revealed that several quasicrystals rank almost first in the magnitude of the coefficient. We then searched the database for the best materials that can be combined with quasicrystals to create high-performance composite thermal diodes. We found that the best choice of material yields a TRR of 3.2, which is higher than the previous highest value for composite thermal diodes.

## II. METHODS

### A. Analysis using Starrydata

Starrydata[20] contained 23,482 temperature dependence curves $\lambda(T)$ of thermal conductivity as of April 4, 2024. From these, we extracted 10,112 curve data that covered the temperature range 320-600 K, i.e., the set of data $\{\lambda(T_1), \lambda(T_2), \ldots, \lambda(T_N)\}$ ($T_1 < T_2 < \ldots < T_N$), satisfying $T_1 < 320$ K and 600 K$< T_N$. For each of the sets of data, we constructed the continuous function $\lambda(T)$ from the discrete data points by connecting adjacent points with a line. Here, for the sets of data with $300\ \mathrm{K} < T_1$, $\lambda(T)$ was extended down to 300 K by a linear extrapolation from $\lambda(T_1)$ and $\lambda(T_2)$. Subsequently, the ranking of the materials with respect to the magnitude of $R = \lambda(600\mathrm{K})/\lambda(300\mathrm{K})$ was determined, and the distribution of this value was examined.

Combinations of materials to construct high-performance composite thermal diodes were searched using the $R$-value ranking of the materials. Here, TRR=$|J_{\mathrm{for}}|/|J_{\mathrm{rev}}|$, which was used to evaluate the performance of the thermal diode, was calculated using analytical formulae based on a simple one-dimensional model, given by Takeuchi et al.[13] Here, $J_{\mathrm{for}}$ in Fig. 1(a) (forward configuration) can be written as follows:

$$J_{\mathrm{for}} = \frac{1}{L}\left(\int_{T_{\mathrm{L}}}^{T_\alpha} \lambda_{\mathrm{B}}(T)dT + \int_{T_\alpha}^{T_{\mathrm{H}}} \lambda_{\mathrm{A}}(T)dT\right), \qquad (1)$$

where $L$ is the total sample length, $T_\alpha$ is the temperature at the interface, and $\lambda_{\mathrm{A}}(T)$ and $\lambda_{\mathrm{B}}(T)$ are thermal conductivities at temperature $T$ for materials A and B, respectively. Similarly, $J_{\mathrm{rev}}$ in Fig. 1(b) (reverse configuration) is given by

$$J_{\mathrm{rev}} = \frac{1}{L}\left(\int_{T_{\mathrm{L}}}^{T_{\beta}} \lambda_{\mathrm{A}}(T)dT + \int_{T_{\beta}}^{T_{\mathrm{H}}} \lambda_{\mathrm{B}}(T)dT\right), \qquad (2)$$

where $T_{\beta}$ is the temperature at the interface. Let the length of materials A and B be $Lx$ and $L(1-x)$, respectively (Figs 1(a) and 1(b)). Then, $x$ is related with $T_{\alpha}$ and $T_{\beta}$ by the equations:

$$\frac{1}{1-x}\int_{T_{\mathrm{L}}}^{T_{\alpha}} \lambda_{\mathrm{B}}(T)dT = \frac{1}{x}\int_{T_{\alpha}}^{T_{\mathrm{H}}} \lambda_{\mathrm{A}}(T)dT \qquad (3)$$

and

$$\frac{1}{x}\int_{T_{\mathrm{L}}}^{T_{\beta}} \lambda_{\mathrm{A}}(T)dT = \frac{1}{1-x}\int_{T_{\beta}}^{T_{\mathrm{H}}} \lambda_{\mathrm{B}}(T)dT \qquad (4)$$

Once $T_{\mathrm{H}}$, $T_{\mathrm{L}}$, $\lambda_{\mathrm{A}}(T)$ and $\lambda_{\mathrm{B}}(T)$ are given, Eqs. (3) and (4) determine $T_{\alpha}$ and $T_{\beta}$ for an arbitrary $x$. Then, Eqs. (1) and (2) determine $J_{\mathrm{for}}$ and $J_{\mathrm{rev}}$, giving TRR=$|J_{\mathrm{for}}|/|J_{\mathrm{rev}}|$. As a result, we obtain TRR as a function of $x$ ( $0 < x < 1$ ), determining its maximum value $\mathrm{TRR}_{\mathrm{max}}$.

## B. Heat transfer simulations

For the thermal diode with the highest $\mathrm{TRR}_{\mathrm{max}}$, heat transfer simulations based on the finite element method were performed using "SOLIDWORKS Simulation" (Dassault Systèmes SolidWorks Corp.) to investigate whether such a high TRR value can be obtained under realistic conditions. Here, we investigated the effects of contact thermal resistance at the interface and heat radiation, which are not considered in the above analytical model[13] of Eqs. (1)-(4).

We modeled a composite thermal diode based on a previous experiment of Takeuchi,[12] as shown in Fig. 2 (a). In this model, materials A and B were cylindrical with a diameter $\phi = 10$ mm and a total length $L$=20 mm, which formed the thermal diode. A pair of copper cylinders with $\phi$=15 mm and $L$=5 mm was attached at both ends of the thermal diode. They were maintained at $T_{\mathrm{H}}$ and $T_{\mathrm{L}}$, and the steady-state distributions of temperature, heat flux, and heat radiation were evaluated. Generally, heat transfer occurs via conduction, convection, or radiation. In all our simulations, convection was neglected on the assumption of the operation of the thermal diode under vacuum, following the experiment by Takeuchi.[12] First, we performed a simulation that only involved conduction with zero thermal resistance at the interface between materials A and B to confirm the validity of the calculations based on the analytical model of Eqs. (1)-(4). The thermal resistance at the interface was then added to the simulation to determine its effect. Finally, simulations considering both conduction and radiation were performed to clarify the effects of radiation. In addition to the setting shown in Fig. 2(a), we performed a simulation using the setting shown in Fig. 2(b), where an aluminum cylinder with a height of 30 mm, inner diameter of 20 mm, and a thickness of 1 mm was placed, acting as a radiation shield. Both surface-to-surface and surface-to-environment radiation were considered. The temperature of the environment was set at 300 K. The emissivities of the copper cylinders, aluminum radiation shield, and thermal diode (materials A and B) were set to 0.1, 0.04, and 0.2, respectively.

# III. RESULTS AND DISCUSSION

## A. Ranking of the positive temperature coefficient of thermal

**conductivity**

Figure 3 shows all 10,112 $\lambda(T)$ curves, where red lines indicate the curves of QCs and lines with other colors indicate those of the others. The values of $\lambda(T)$ are distributed roughly in the range of $10^{-1} - 10^{3}\ \mathrm{W \cdot m^{-1} \cdot K^{-1}}$. Eighteen QC curves were extracted and are plotted in Fig. 4(a). The $\lambda(300\mathrm{K})$ values of them are in the range of $6 \times 10^{-1} - 4\ \mathrm{W \cdot m^{-1} \cdot K^{-1}}$. The $\lambda(T)$ of all QCs show a positive temperature coefficient and almost linear dependence in the log-plot. This indicates that $\lambda(T)$ increases almost exponentially with increasing temperature, which is very rare.

Figure 5 presents the distribution of $R = \lambda(600\mathrm{K})/\lambda(300\mathrm{K})$, where arrows indicate the $R$ values of QCs. The number of curves with $R < 1$, i.e., $\lambda(600\mathrm{K}) < \lambda(300\mathrm{K})$, is much larger than that with $R > 1$, i.e., $\lambda(600\mathrm{K}) > \lambda(300\mathrm{K})$: the former is 8,375 (82.8 %) and the latter 1,737 (17.2 %). Curves with $R > 2$ were rare ($n$=71, 0.7 %). These included 13 QC curves. The ranking of $R$ after removing erroneous or highly unreliable data are listed in Table 1. The removed data included electronic (i.e., not total) thermal conductivity data, data extrapolated from measured data below 200 K, and data that were largely different between heating and cooling. Note that this list includes several data that do not reflect the bulk properties of a single material, such as those of nanosheet samples, nanocomposite samples, granular samples of different ceramics particles, and nanostructured ceramics, as indicated in red in Table 1. Excluding these, 12 of the 31 were QCs of various alloy systems: Al-Cu-Fe, Al-Cu-Pt-Fe, Al-Cu-Au-Fe, Au-Al-Tm, Al-Pd-Re, Al-Pd-Re-Ru, and Al-Pd-Ru. The $\lambda(T)$ curves of the 19 data indicated in black in Table 1 are shown in Fig. 4(b).

The above results in Table 1 and Fig. 5 verify that QCs are in the top group among all materials in terms of the magnitude of the positive temperature coefficient of thermal conductivity above room temperature. So far, several physical properties characteristic of QCs have been identified, including unusually low electrical conductivity for a metal, a large positive temperature coefficient of electrical conductivity, and hard and brittle mechanical properties at room temperature. However, QCs do not belong to the top group among all materials in these properties. Therefore, the exceptionally high positive temperature coefficient of thermal conductivity above room temperature is the only physical property in which QCs rank almost first among all materials and is the most distinct property of QCs.

The origin of this property has been discussed in relation to the pseudogap in the electronic density of states,[11,12] often seen for QC alloys. In fact, when the energy dependence of the electronic density of states $D(E)$ has a deep pseudogap and the chemical potential $\mu$ is located near the bottom of the pseudogap, electronic thermal conductivity $\lambda_{\mathrm{e}}(T)$ can exhibit a large positive temperature coefficient. When $\mu$ is close to the energy at the bottom of the pseudogap, $\lambda_{\mathrm{e}}(T)$ can be written as

$$\lambda_{\mathrm{e}}(T) = \frac{1}{e^2 T}\int_{-\infty}^{\infty} \sigma_{\mathrm{s}}(E,T)(E-\mu)^2\left(-\frac{\partial f(E,T)}{\partial E}\right) dE\,, \qquad (5)$$

where $e$ is the unit charge of electron, $\sigma_{\mathrm{s}}(E,T)$ is the spectral conductivity, and $f(E,T)$ is the Fermi-Dirac distribution function. Adopting the relaxation time approximation with the isotropic electronic structure, we can write $\sigma_{\mathrm{s}}(E,T)$ as

$$\sigma_{\mathrm{s}}(E,T) = \frac{e^2}{3} D(E) v_G^2(E)\tau(E,T)\,, \qquad (6)$$

where $v_G(E)$ and $\tau(E,T)$ are the group velocity and relaxation time, respectively. Neglecting the energy dependence of $v_G(E)$ and $\tau(E,T)$ and inserting Eq. (6) into (5),

we obtain

$$\lambda_{\mathrm{e}}(T) = \frac{v_G^2}{3}\frac{1}{T}\tau(T)\int_{-\infty}^{\infty} D(E)\,(E-\mu)^2\left(-\frac{\partial f(E,T)}{\partial E}\right)dE$$

$$= \frac{v_G^2}{3}\frac{1}{T}\tau(T)F_1(T), \qquad (7)$$

where

$$W_1(E,T) = (E-\mu)^2\left(-\frac{\partial f(E,T)}{\partial E}\right) \qquad (8)$$

is a window function for $D(E)$ in $F_1(T)$. In Eq. (7), both $1/T$ and $\tau(T)$ are decreasing functions of $T$ but $F_1(T)$ can be an increasing function of $T$ with a large coefficient, when $\mu$ is at the bottom of a deep pseudogap of $D(E)$, as shown in Fig. 6(a). Here, the window function $W_1(E,T_0)$ for a given $T = T_0$ has two peaks at $E = \mu \pm 2.4k_{\mathrm{B}}T_0$ ($k_{\mathrm{B}}$: the Boltzmann constant), the separation of which increases with increasing temperature, as shown in Fig. 6(b). This should result in a significant increase of $F_1(T)$ as $T$ increases, resulting in an increase function of $\lambda_{\mathrm{e}}(T)$.

On the other hand, the electrical conductivity $\sigma(T)$ is given by

$$\sigma(T) = \int_{-\infty}^{\infty}\sigma_{\mathrm{s}}(E,T)\left(-\frac{\partial f(E,T)}{\partial E}\right)dE$$

$$= \frac{e^2v_G^2}{3}\tau(T)\int_{-\infty}^{\infty}D(E)\left(-\frac{\partial f(E,T)}{\partial E}\right)dE,$$

$$= \frac{e^2v_G^2}{3}\tau(T)F_2(T), \qquad (9)$$

where

$$W_2(E,T) = -\frac{\partial f(E,T)}{\partial E} \qquad (10)$$

is a window function for $D(E)$ in $F_2(T)$. The window function $W_2(E,T_0)$ for a given

$T = T_0$ has a single peak at $E = \mu$, whose width increases with increasing temperature, as shown in Fig. 6(c). This indicates that the function $F_2(T)$ is an increasing function of $T$ for the electronic density of states with a pseudogap such as $D(E)$ in Fig. 6(a). If the positive temperature coefficient in $F_2(T)$ overcomes the negative temperature coefficient in $\tau(T)$ in Eq. (9), $\sigma(T)$ becomes an increasing function of $T$.

Among the twelve papers of the $\lambda(T)$ data of QCs in Table 1, ten papers have also reported $\sigma(T)$ data. From the figures of $\sigma(T)$ in the papers, we estimated $R_\sigma = \sigma(600\text{K})/\sigma(300\text{K})$ values, which are listed in Tabel 2, together with the $R = \lambda(600\text{K})/\lambda(300\text{K})$ values. For all the QCs, $R_\sigma > 1$, i.e., $\sigma(600\text{K}) > \sigma(300\text{K})$. The $R_\sigma$ values are $1.7 - 2.0$, except for $Au_{49}Al_{34}Tm_{17}$ (1.1) and $Al_{71}Pd_{20}Ru_9$ (1.2). No appreciable correlation can be seen between $R$ and $R_\sigma$, although some correlation would be expected in the above theoretical framework. More detailed analysis is needed to resolve this discrepancy.

## B. Search for high-performance thermal diodes

The data with $R < 0.3$, after removing erroneous or highly unreliable data, are listed in Table 3, and the $\lambda(T)$ curves of them are shown in Fig. 4(c). For all combinations of the black and blue samples in Table 1 as material A and the samples in Table 3 as material B, we calculated the $TRR_{max}$ for ($T_H$, $T_L$)=(600 K, 300 K) and (700 K, 300 K) according to the analytical model[13] of Eqs. (1)-(4) in Sec. IIA. For $Ga_{0.05}Sn_{0.95}Bi_2Te_4$, $Ag_2Se$, $SnBi_4Te_7$, $In_{0.05}Sn_{0.95}Bi_2Te_4$, $In_{0.05}SnBi_{1.95}Te_4$, and $Ge_{12}In_2Te_{15}$, the data ended at a temperature between 600 and 700K. For these, linear extrapolation was performed from the two data points closest and second-closest to 700 K to extend $\lambda(T)$ up to 700 K. In Table 4, the rankings of $TRR_{max}$ for ($T_H$, $T_L$)=(600 K, 300 K) are listed together with the values of $TRR_{max}$ for ($T_H$, $T_L$)=(700 K, 300 K). Here, $X$ indicates the $x$ value that yields $TRR_{max}$ for ($T_H$, $T_L$)=(600 K, 300 K). Down to rank 25, material B consists exclusively of CuAgSe or (Zn, Ni, Co)-doped CuAgSe. Materials other than (doped) CuAgSe appeared for the first time at rank 69, which was $MoTe_{1.97}I_{0.03}$. The best combinations without (doped) CuAgSe are listed at the bottom of Table 3. The combination of QC-$Al_{61.5}Cu_{26.5}Fe_{12}$ and $CuGaTe_2$ at rank 91 was previously reported by Takeuchi.[12] This gave TRR=2.2 by experiment and 2.26 by calculation for ($T_H$, $T_L$)=(900 K, 300 K), which is the highest ever for a composite thermal diode containing QC as a component. Starrydata contained several $\lambda(T)$ curves of $CuGaTe_2$, reported by different groups. The curve of $CuGaTe_2$ at rank 91 in Table 3 is that reported by Ye et al.[41] We calculated the TRR value for ($T_H$, $T_L$)=(700 K, 300 K) using $\lambda(T)$ of $CuGaTe_2$ reported by Takeuchi,[12] and obtained TRR=1.83, which is slightly smaller than TRR=2.22 in Table 3. In any case, TRR=3.18 for ($T_H$, $T_L$)=(700 K, 300 K) at rank 1 for QC-$Al_{61.5}Cu_{26.5}Fe_{12}$ and CuAgSe in Table 3 is much higher than the previous highest value.

To the best of our knowledge, the best composite thermal diode to date, irrespective of containing a QC, was that consisting of $Ag_2S_{0.6}Se_{0.4}$ as material A and $Ag_2S_{0.1}Te_{0.9}$ as material B, which showed TRR=2.7 for ($T_H$, $T_L$)=(413 K, 300 K).[19] Here, both materials A and B exhibit rapid changes in thermal conductivity due to phase transitions, resulting in a high TRR value for relatively small temperature difference $\Delta T = T_H - T_L$. Although the value of $\Delta T$ is larger than this, TRR=3.18 obtained in the present study is higher than the previous highest value.

The temperature dependences of the thermal conductivities of the two materials at rank 1 in Table 4 are shown in Fig. 7(a). With increasing temperature, CuAgSe exhibits a sharp drop in thermal conductivity at about 450 K, which is due to a phase transition from the low-temperature semi-metallic phase to the high-temperature semiconduction phase.[39] In Fig. 7(b), TRR is calculated as a function of $x$ for ($T_{\mathrm{H}}$, $T_{\mathrm{L}}$)=(600 K, 300 K) and (700 K, 300 K) using Eqs. (1)-(4) in Sec. IIA. In Fig. 7(b), $\mathrm{TRR_{max}}$ is 2.58 for ($T_{\mathrm{H}}$, $T_{\mathrm{L}}$)=(600 K, 300 K) and 3.18 for (700 K, 300 K).

## C. Heat transfer simulations

In this subsection, we present the results of the heat transfer simulations for the best material combination (material A: $Al_{61.5}Cu_{26.5}Fe_{12}$; material B: CuAgSe) with ($T_{\mathrm{H}}$, $T_{\mathrm{L}}$)=(700 K, 300 K).

First, we present the results of the simulation, in which heat transfer was assumed to occur only through conduction. Here, in principle, the average heat flux at a cross section of the cylindrical diode in the steady state is constant throughout the length of the diode from the high-temperature end to the low-temperature end. In Figs. 8(a) and 8(b), the simulation results (solid circles) for the heat flux $J_{\mathrm{for}}$ in the forward configuration, $J_{\mathrm{rev}}$ in the reverse configuration, and TRR=$|J_{\mathrm{for}}|/|J_{\mathrm{rev}}|$ are compared with those of the analytical calculations (lines) using Eqs. (1)-(4) in Sec. IIA. The simulation results agreed well with the calculation results, indicating the validity of the analytical model of Eqs. (1)-(4).

Next, we investigated the effect of thermal resistance at the interface between materials A and B, where heat transfer was assumed to occur only thorough conduction. In Figs. 9(a) and 9(b), $J_{\mathrm{for}}$, $J_{\mathrm{rev}}$, and TRR=$|J_{\mathrm{for}}|/|J_{\mathrm{rev}}|$ for $x = 0.7$ are plotted against the thermal resistance $R_t$. The effect of the thermal resistance is negligible for $R_t \leq 10^{-3}\mathrm{K} \cdot \mathrm{m}^2 \cdot \mathrm{W}^{-1}$. For $R_t = 10^{-2}\mathrm{K} \cdot \mathrm{m}^2 \cdot \mathrm{W}^{-1}$, a significant change can be observed in the heat flux, and the TRR is largely reduced. Practically, $R_t \leq 10^{-4} \mathrm{K} \cdot \mathrm{m}^2 \cdot \mathrm{W}^{-1}$ is easily realizable, e.g., using thermal grease, and therefore, the effect of the thermal resistance at the interface can be neglected in the use of the thermal diode.

Finally, simulations were conducted to examine the effects of radiation. In the forward configuration of Fig. 2(a) with $x = 0.7$, the heat inflow from the copper cylinder at $T_\mathrm{H} = 700$ K to material A ($Al_{61.5}Cu_{26.5}Fe_{12}$) was $Q_\mathrm{for}^\mathrm{in}$ =6.35 W, whereas the heat outflow from material B (CuAgSe) to the copper cylinder at $T_\mathrm{L} = 300$ K was $Q_\mathrm{for}^\mathrm{out}$ =5.66 W, as in Table 5. Then, the net radiation from the side of the diode was calculated to be $Q_\mathrm{for}^\mathrm{rad} = Q_\mathrm{for}^\mathrm{in} - Q_\mathrm{for}^\mathrm{out}$ =0.69 W, leading to a radiation loss of $Q_\mathrm{for}^\mathrm{rad}/Q_\mathrm{for}^\mathrm{in}$ =0.108. The corresponding values for the reverse configuration are listed in Table 5. The heat flows $Q_\mathrm{for}$ and $Q_\mathrm{rev}$ at $x = 0.7$ under the no radiation condition were calculated to be 5.80 and 1.82 W, respectively, from the heat fluxes shown in Fig. 8(a). These values are comparable to the corresponding outflow values but considerably smaller than the inflow values. If we estimate the TRR values using the heat inflows and outflows, we obtain $Q_\mathrm{for}^\mathrm{in}/Q_\mathrm{rev}^\mathrm{in} = 3.13$ and $Q_\mathrm{for}^\mathrm{out}/Q_\mathrm{rev}^\mathrm{out} = 3.25$, respectively. In the configuration shown in Fig. 2(b), that is, with the aluminum radiation shield, the obtained heat flow values are listed in Table 5. The radiation loss was reduced from 0.108 to 0.0566 in the forward configuration and from 0.140 to 0.0394 in the reverse configuration, confirming that the radiation shield worked effectively.

In summary of this subsection, the composite thermal diode was found to operate with negligible radiation loss, achieving TRR≈3.2 under realistic conditions.

## IV. CONCLUSIONS

We analyzed the large physical property database "Starrydata" for the following two purposes: (1) to determine the degree of QC specificity with respect to the large positive temperature coefficient of thermal conductivity in the temperature range above room temperature, and (2) to search for optimal materials to combine with QCs to construct high performance composite thermal diodes.

For (1), we investigated the ranking of the ratio of thermal conductivities $R = \lambda(600\mathrm{K})/\lambda(300\mathrm{K})$ for more than 10,000 $\lambda(T)$ curves in the database. The number of curves with $R > 2$ was 71 (0.7 %), including 13 QC curves. Notably, several QCs ranked almost first in the $R$ value. No other physical property is such that QCs belong to the top group of all materials, and therefore, the exceptionally high positive temperature coefficient of thermal conductivity above room temperature is the most distinct property of QCs. The physical origin of this property was discussed in relation to the pseudogap in the electronic density of states. Some correlation was expected between the magnitude of the temperature coefficient of thermal conductivity and that of electronic conductivity, but it was not detected. Further analysis is needed to resolve this issue.

For (2), we extracted the samples with $R < 0.3$, and combined them with the samples with $R > 2.18$ to construct a composite thermal diode. Based on the analytical model, we calculated the thermal rectification ratio (TRR) for each combination and obtained a ranking of the TRR values. The combination of QC-$Al_{61.5}Cu_{26.5}Fe_{12}$ and CuAgSe ranked first, with TRR=2.6 for ($T_\mathrm{H}$, $T_\mathrm{L}$)=(600 K, 300 K) and 3.2 for (700 K, 300 K). The latter value was the highest among those reported to date for composite thermal diodes. For this diode, heat transfer simulations based on the finite element method were performed to evaluate its performance under realistic conditions. It was confirmed that TRR≈3.2 is indeed realizable under realistic conditions. The effect of thermal resistance at the material interface was found to be practically negligible. The radiation shield was found to be effective in significantly reducing radiation losses.

## ACKNOWLEDGEMENTS

We thank R. Yoshida (ISM), R. Tamura (TUS), K. Kimura (NIMS, ISM), and E. Fujita (NIMS) for fruitful discussion and comments. This study was supported by a JST CREST Grant (No. JPMJCR22O3), Japan, and KAKENHI Grant-in-Aid (No. JP19H05821) from JSPS.

## AUTHOR DECLARATIONS

### Conflict of Interest

The authors have no conflicts to disclose.

### Author Contributions

**Takashi Kurono**: Formal analysis (lead); Investigation (lead); Writing – review & editing (equal). **Jijia Zhang**: Formal analysis (equal); Investigation (equal); Writing – review & editing (equal). **Yasushi Kamimura**: Formal analysis (equal); Investigation (equal); Writing – review & editing (equal). **Keiichi Edagawa**: Conceptualization (lead); Formal analysis (equal); Funding acquisition (lead); Investigation (equal); Resources (lead); Supervision (lead); Writing – original draft (lead); Writing – review & editing (equal).

## DATA AVAILABILITY

The data that support the findings of this study are available from the corresponding author upon reasonable request.

TABLE 1. Ranking of the ratio of the thermal conductivity $R = \lambda(600\text{K})/\lambda(300\text{K})$. "QC" and "1/1-AC" in Note indicate quasicrystal and 1/1 approximant crystal, respectively. The data not reflecting a bulk property of a single material, those of QCs, and others are shown in red, blue, and black, respectively. The sample composition of the QC at rank 33 was $Al_{62}Cu_{25.5}Fe_{23.5}$ in Starrydata, but in this Table it is corrected to $Al_{62}Cu_{25.5}Fe_{12.5}$ in accordance with the original paper.[35]

| Ranking | Sample | $R$ | Note | Reference |
|---|---|---|---|---|
| 1 | $Al_{0.01}(CuO)_{0.99}$ | 13.38 | nano-sheet | 21 |
| 2 | $Al_{0.03}(CuO)_{0.97}$ | 10.66 | nano-sheet | 21 |
| 3 | $Al_{0.005}(CuO)_{0.995}$ | 10.01 | nano-sheet | 21 |
| 4 | $Al_{0.05}(CuO)_{0.95}$ | 7.62 | nano-sheet | 21 |
| 5 | $(SiO_2)_x(Al_2O_3)_y$ | 6.00 | granular ceramics particles | 22 |
| 6 | CuO nano-sheet | 5.55 | nano-sheet | 21 |
| 7 | $Ga_{0.05}Sn_{0.95}Bi_2Te_4$ | 3.86 | | 23 |
| 8 | $Al_{0.02}(ZnO)_{0.98}$ | 3.40 | nano-structured ceramics | 24 |
| 9 | $La_{0.067}Sr_{0.9}TiO_3C_{0.09}$ | 3.30 | nano-composite | 25 |
| 10 | $Al_{61.5}Cu_{26.5}Fe_{12}$ | 3.19 | QC | 6 |
| 11 | $Al_{63.2}Cu_{24.2}Pt_{0.9}Fe_{11.3}$ | 2.84 | QC | 26 |
| 12 | $Al_{62}Cu_{25.5}Fe_{12.5}$ | 2.82 | QC | 6 |
| 13 | $Au_{49}Al_{36}Tm_{15}$ | 2.80 | 1/1-AC | 7 |
| 14 | $SnBi_4Te_7$ | 2.78 | | 27 |
| 15 | $Ag_2Se$ | 2.76 | phase transition | 28 |
| 16 | $La_{0.9}Sr_{0.1}CoO_3$ | 2.71 | | 29 |
| 17 | $La_{0.9}Sr_{0.1}CoO_3$ | 2.60 | | 30 |
| 18 | $Al_{63.2}Cu_{23.8}Pt_{1.7}Fe_{11.3}$ | 2.56 | QC | 26 |
| 19 | $Au_{49}Al_{36}Tm_{15}$ | 2.55 | 1/1-AC | 7 |
| 20 | $Al_{63.1}Cu_{24.2}Au_{0.9}Fe_{11.8}$ | 2.52 | QC | 26 |
| 21 | $Al_{63.1}Cu_{22.3}Pt_{2.6}Fe_{12}$ | 2.49 | QC | 26 |
| 22 | $Na_{0.4}Co_{0.2}Ti_{0.8}O_2$ | 2.45 | | 31 |
| 22 | $Al_{63}Cu_{25}Fe_{12}$ | 2.45 | QC | 26 |
| 24 | $Ge_{12}In_2Te_{15}$ | 2.44 | | 32 |
| 24 | $Al_{0.01}(ZnO)_{0.99}$ | 2.44 | nano-structured ceramics | 24 |
| 26 | $La_{0.95}Sr_{0.05}Co_{O3}$ | 2.36 | | 30 |
| 27 | $CuOC_{0.07}$ | 2.34 | nano-sheet | 33 |
| 28 | $Au_{49}Al_{34}Tm_{17}$ | 2.31 | QC | 7 |
| 28 | $In_{0.05}SnBi_{1.95}Te_4$ | 2.31 | | 23 |
| 30 | $Cu_{12}Sb_4S_{13}$ | 2.30 | | 34 |
| 31 | $(Nd_{0.8}Ce_{0.2})_2Zr_2O_{7.2}$ | 2.29 | | 35 |
| 32 | $In_{0.05}Sn_{0.95}Bi_2Te_4$ | 2.28 | | 23 |
| 33 | $Al_{62}Cu_{25.5}Fe_{12.5}$ | 2.26 | QC | 36 |
| 33 | $Ag_{2.0027}Se$ | 2.26 | phase transition | 28 |
| 35 | $Al_{68.5}Pd_{22.9}Re_{8.6}$ | 2.25 | QC | 8 |
| 36 | $Al_{71}Pd_{20}Re_{2.7}Ru_{6.3}$ | 2.24 | QC | 9 |
| 37 | $Cu_{0.07}Bi_2Te_3$ | 2.23 | single-crystal z-direction | 37 |
| 38 | $Al_{71}Pd_{20}Ru_9$ | 2.18 | QC | 9 |
| 38 | $Cu_{0.07}Bi_2Te_3$ | 2.18 | single-crystal z-direction | 37 |

TABLE 2. The values of $R = \lambda(600\mathrm{K})/\lambda(300\mathrm{K})$ and $R_\sigma = \sigma(600\mathrm{K})/\sigma(300\mathrm{K})$ for ten QC samples. The sample composition of the QC of ref. 35 was $Al_{62}Cu_{25.5}Fe_{23.5}$ in Starrydata, but in this Table it is corrected to $Al_{62}Cu_{25.5}Fe_{12.5}$ in accordance with the original paper.

| Sample | $R$ | $R_\sigma$ | Reference |
|---|---|---|---|
| $Al_{63.2}Cu_{24.2}Pt_{0.9}Fe_{11.3}$ | 2.84 | 1.7 | 26 |
| $Al_{63.2}Cu_{23.8}Pt_{1.7}Fe_{11.3}$ | 2.56 | 1.9 | 26 |
| $Al_{63.1}Cu_{24.2}Au_{0.9}Fe_{11.8}$ | 2.52 | 1.8 | 26 |
| $Al_{63.1}Cu_{22.3}Pt_{2.6}Fe_{12}$ | 2.49 | 1.8 | 26 |
| $Al_{63}Cu_{25}Fe_{12}$ | 2.45 | 1.9 | 26 |
| $Au_{49}Al_{34}Tm_{17}$ | 2.31 | 1.1 | 7 |
| $Al_{62}Cu_{25.5}Fe_{12.5}$ | 2.26 | 1.8 | 36 |
| $Al_{68.5}Pd_{22.9}Re_{8.6}$ | 2.25 | 2.0 | 8 |
| $Al_{71}Pd_{20}Re_{2.7}Ru_{6.3}$ | 2.24 | 1.7 | 9 |
| $Al_{71}Pd_{20}Ru_{9}$ | 2.18 | 1.2 | 9 |

TABLE 3. Ranking of the ratio of the thermal conductivity $R = \lambda(600\mathrm{K})/\lambda(300\mathrm{K})$. The samples with $R < 0.3$ are listed.

| Ranking | Sample | $R$ | Reference |
|---|---|---|---|
| 1 | CuAgSe | 0.16 | 38 |
| 2 | $CuAg_{0.05}Ga_{0.95}Te_2$ | 0.18 | 39 |
| 3 | $Zn_{0.02}CuAgSe$ | 0.21 | 38 |
| 4 | $Ni_{0.02}CuAgSe$ | 0.23 | 38 |
| 4 | $Co_{0.02}CuAgSe$ | 0.23 | 38 |
| 6 | $MoTe_{1.96}I_{0.04}$ | 0.24 | 40 |
| 6 | $MoTe_{1.97}I_{0.03}$ | 0.24 | 40 |
| 6 | $CuGaTe_2$ | 0.24 | 41 |
| 9 | $In_2Se_3$ | 0.25 | 42 |
| 9 | $Cu_{1.8}Sn_{0.1}Mn_{0.1}S$ | 0.25 | 43 |
| 11 | $Pb_{0.9735}TeK_{0.0125}Na_{0.014}$ | 0.26 | 44 |
| 12 | $MoTe_{1.98}I_{0.02}$ | 0.27 | 40 |
| 12 | $Fe_2O_3$ | 0.27 | 45 |
| 12 | $In_2O_3$ | 0.27 | 46 |
| 15 | $CdTe_{0.995}Cl_{0.005}$ | 0.28 | 47 |
| 16 | $MoTe_{1.99}I_{0.01}$ | 0.29 | 40 |
| 16 | $Zn_{0.98}Al_{0.02}O$ | 0.29 | 48 |
| 16 | BeO | 0.29 | 49 |
| 16 | $Cu_2CdSnSe_4$ | 0.29 | 50 |
| 16 | $Cu_2ZnGeS_4$ | 0.29 | 51 |
| 21 | $Cu_2ZnGeSe_4$ | 0.30 | 52 |
| 21 | $Cu_2ZnGeSe_4$ | 0.30 | 53 |
| 21 | $CuInTe_2$ | 0.30 | 54 |
| 21 | $CuGa_{0.98}Gd_{0.02}Te_2$ | 0.30 | 55 |

TABLE 4. Ranking of $TRR_{max}$ for $(T_H, T_L)$=(600 K, 300 K). $R$ denotes the ratio of the thermal conductivity $\lambda(600K)/\lambda(300K)$. "QC" and "1/1-AC" in Note indicate quasicrystal and 1/1 approximant crystal, respectively. $X$ indicates the $x$ value that gives $TRR_{max}$ for $(T_H, T_L)$=(600 K, 300 K). $TRR_{max}$ values for $(T_H, T_L)$=(700 K, 300 K) are also shown in the last column.

| Ranking | Material A | | | Material B | | $X$ | $TRR_{max}$ | $TRR_{max}$(700K) |
|---|---|---|---|---|---|---|---|---|
| | Composition | $R$ | Note | Composition | $R$ | | | |
| 1 | $Al_{61.5}Cu_{26.5}Fe_{12}$ | 3.19 | QC | CuAgSe | 0.16 | 0.62 | 2.58 | 3.18 |
| 2 | $Al_{63.2}Cu_{24.2}Pt_{0.9}Fe_{11.3}$ | 2.84 | QC | CuAgSe | 0.16 | 0.57 | 2.47 | 3.15 |
| 3 | $Au_{49}Al_{36}Tm_{15}$ | 2.80 | 1/1-AC | CuAgSe | 0.16 | 0.78 | 2.42 | 2.95 |
| 4 | $Al_{62}Cu_{25.5}Fe_{12.5}$ | 2.82 | QC | CuAgSe | 0.16 | 0.53 | 2.41 | 3.01 |
| 5 | $Ga_{0.05}Sn_{0.95}Bi_2Te_4$ | 3.86 | | CuAgSe | 0.16 | 0.39 | 2.38 | 2.94 |
| 6 | $Al_{61.5}Cu_{26.5}Fe_{12}$ | 3.19 | QC | $Ni_{0.02}CuAgSe$ | 0.23 | 0.64 | 2.36 | 2.79 |
| 7 | $Al_{63.2}Cu_{23.8}Pt_{1.7}Fe_{11.3}$ | 2.56 | QC | CuAgSe | 0.16 | 0.56 | 2.35 | 3.01 |
| 8 | $Ag_2Se$ | 2.76 | phase transition | CuAgSe | 0.16 | 0.66 | 2.34 | 2.53 |
| 9 | $Au_{49}Al_{36}Tm_{15}$ | 2.55 | 1/1-AC | CuAgSe | 0.16 | 0.78 | 2.33 | 2.88 |
| 10 | $Al_{61.5}Cu_{26.5}Fe_{12}$ | 3.19 | QC | $Zn_{0.02}CuAgSe$ | 0.21 | 0.65 | 2.32 | 2.83 |
| 11 | $SnBi_4Te_7$ | 2.78 | | CuAgSe | 0.16 | 0.34 | 2.31 | 2.89 |
| 12 | $Al_{63.1}Cu_{24.2}Au_{0.9}Fe_{11.8}$ | 2.52 | QC | CuAgSe | 0.16 | 0.57 | 2.31 | 2.94 |
| 13 | $Al_{63.1}Cu_{22.3}Pt_{2.6}Fe_{12}$ | 2.49 | QC | CuAgSe | 0.16 | 0.55 | 2.29 | 2.92 |
| 14 | $Al_{61.5}Cu_{26.5}Fe_{12}$ | 3.19 | QC | $Co_{0.02}CuAgSe$ | 0.23 | 0.63 | 2.28 | 2.75 |
| 15 | $Al_{63.2}Cu_{24.2}Pt_{0.9}Fe_{11.3}$ | 2.84 | QC | $Ni_{0.02}CuAgSe$ | 0.23 | 0.58 | 2.27 | 2.78 |
| 16 | $Al_{63}Cu_{25}Fe_{12}$ | 2.45 | QC | CuAgSe | 0.16 | 0.57 | 2.27 | 2.92 |
| 17 | $In_{0.05}Sn_{0.95}Bi_2Te_4$ | 2.28 | | CuAgSe | 0.16 | 0.39 | 2.26 | 2.88 |
| 18 | $La_{0.9}Sr_{0.1}CoO_3$ | 2.71 | | CuAgSe | 0.16 | 0.63 | 2.24 | 2.59 |
| 19 | $In_{0.05}SnBi_{1.95}Te_4$ | 2.31 | | CuAgSe | 0.16 | 0.38 | 2.24 | 2.77 |
| 20 | $La_{0.95}Sr_{0.05}CoO_3$ | 2.36 | | CuAgSe | 0.16 | 0.62 | 2.23 | 2.67 |
| 21 | $Al_{63.2}Cu_{24.2}Pt_{0.9}Fe_{11.3}$ | 2.84 | QC | $Zn_{0.02}CuAgSe$ | 0.21 | 0.59 | 2.23 | 2.81 |
| 22 | $Ge_{12}In_2Te_{15}$ | 2.44 | | CuAgSe | 0.16 | 0.52 | 2.23 | 2.74 |
| 23 | $Al_{62}Cu_{25.5}Fe_{12.5}$ | 2.26 | QC | CuAgSe | 0.16 | 0.59 | 2.23 | 2.76 |
| 24 | $Au_{49}Al_{34}Tm_{17}$ | 2.31 | QC | CuAgSe | 0.16 | 0.78 | 2.22 | 2.83 |
| 25 | $Au_{49}Al_{36}Tm_{15}$ | 2.80 | 1/1-AC | $Ni_{0.02}CuAgSe$ | 0.23 | 0.79 | 2.22 | 2.60 |
| | | | | | | | | |
| 69 | $Al_{61.5}Cu_{26.5}Fe_{12}$ | 3.19 | QC | $MoTe_{1.97}I_{0.03}$ | 0.24 | 0.57 | 2.04 | 2.12 |
| 70 | $Al_{61.5}Cu_{26.5}Fe_{12}$ | 3.19 | QC | $Zn_{0.98}Al_{0.02}O$ | 0.24 | 0.19 | 2.04 | 2.14 |
| 73 | $Al_{61.5}Cu_{26.5}Fe_{12}$ | 3.19 | QC | $MoTe_{1.96}I_{0.04}$ | 0.24 | 0.54 | 2.04 | 2.16 |
| 91 | $Al_{61.5}Cu_{26.5}Fe_{12}$ | 3.19 | QC | $CuGaTe_2$ | 0.24 | 0.42 | 2.00 | 2.22 |

TABLE 5. The heat inflows $Q_{\mathrm{for}}^{\mathrm{in}}$ and $Q_{\mathrm{rev}}^{\mathrm{in}}$ and outflows $Q_{\mathrm{for}}^{\mathrm{out}}$ and $Q_{\mathrm{rev}}^{\mathrm{out}}$, net radiations $Q_{\mathrm{for}}^{\mathrm{rad}}$ and $Q_{\mathrm{rev}}^{\mathrm{rad}}$, and radiation losses $Q_{\mathrm{for}}^{\mathrm{rad}}/Q_{\mathrm{for}}^{\mathrm{in}}$ and $Q_{\mathrm{rev}}^{\mathrm{rad}}/Q_{\mathrm{rev}}^{\mathrm{in}}$ in the forward and reverse configurations with and without aluminum radiation shield.

| Shield | X | $Q_{\mathrm{X}}^{\mathrm{in}}$ (W) | $Q_{\mathrm{X}}^{\mathrm{out}}$ (W) | $Q_{\mathrm{X}}^{\mathrm{rad}}$ (W) | $Q_{\mathrm{X}}^{\mathrm{rad}}/Q_{\mathrm{X}}^{\mathrm{in}}$ |
|---|---|---|---|---|---|
| Without shield | for | 6.35 | 5.66 | 0.69 | 0.108 |
| | rev | 2.03 | 1.74 | 0.29 | 0.140 |
| With shield | for | 6.12 | 5.78 | 0.34 | 0.0566 |
| | rev | 1.93 | 1.86 | 0.07 | 0.0394 |

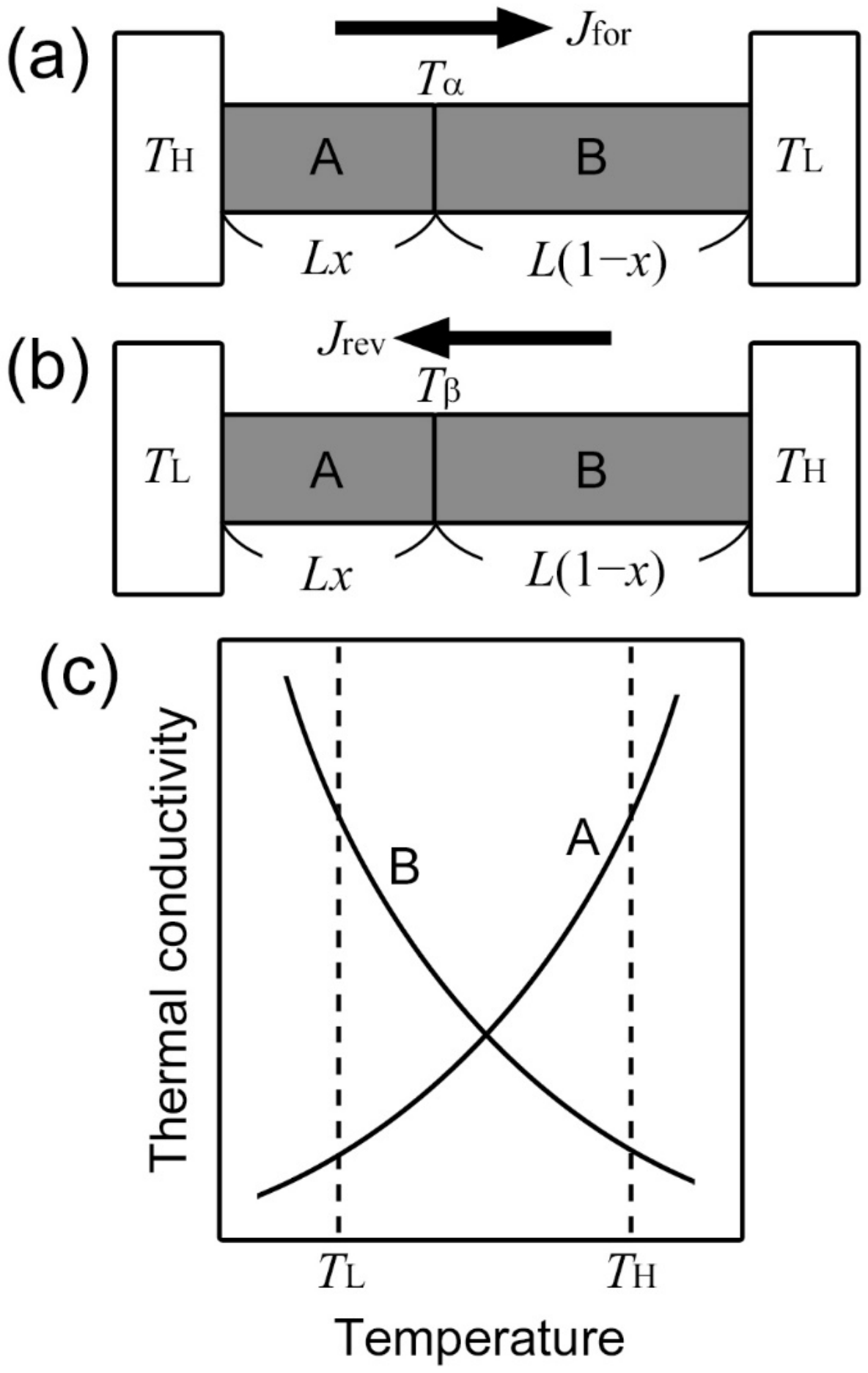

FIG. 1. Schematics of the structure of the composite thermal diode comprising materials A and B[17,18] (a) Forward configuration. (b) Reverse configuration. Both ends of the diode are connected to thermal reservoirs with temperatures of $T_H$ and $T_L$ ($T_H > T_L$), respectively. (c) Temperature dependences of the thermal conductivity for materials A and B.

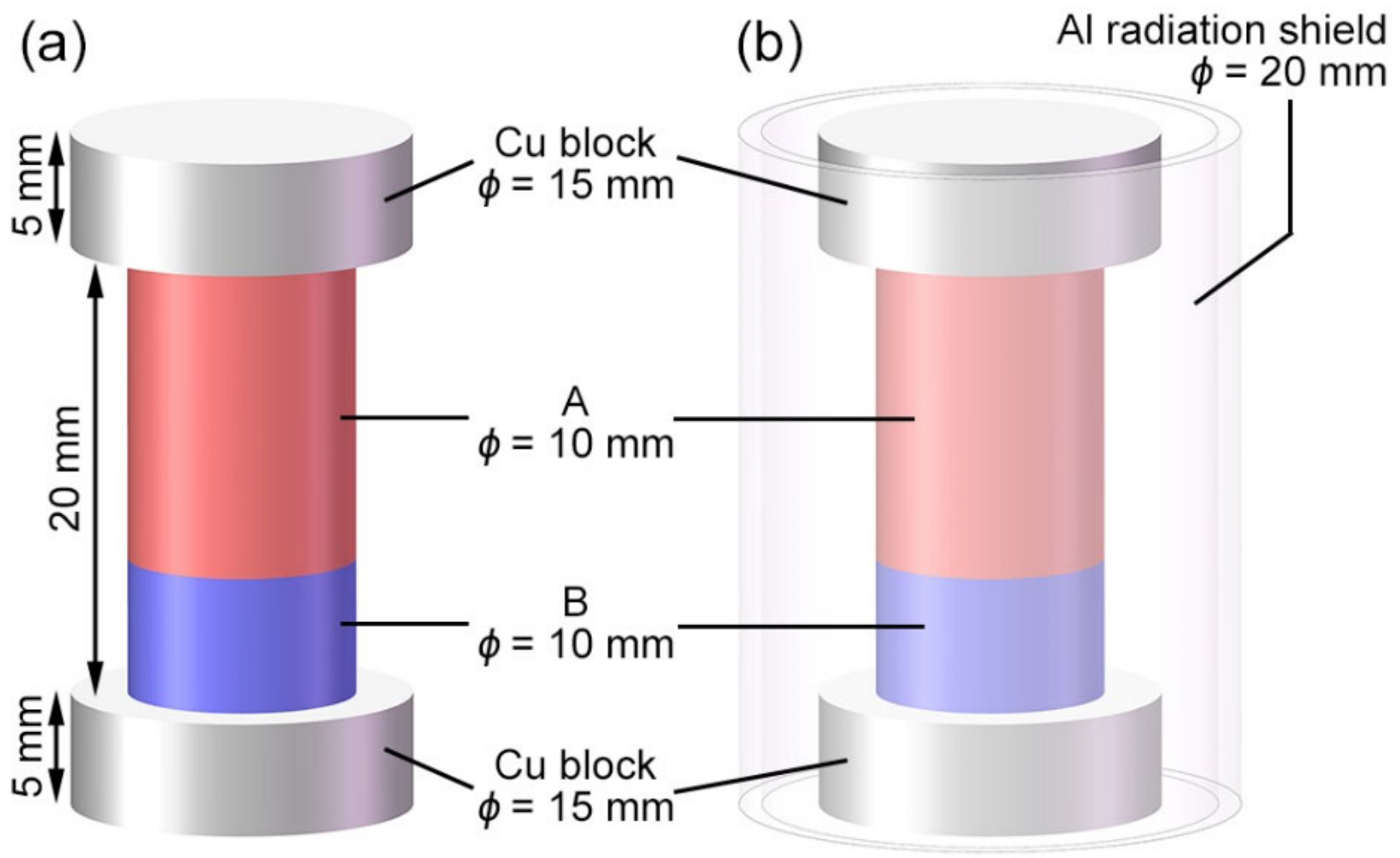

FIG. 2. The settings used for heat transfer simulations. A composite thermal diode consisting of materials A and B was sandwiched by copper cylinders. The structure not covered with a cylindrical aluminum radiation shield (a) and that covered with a cylindrical aluminum radiation shield (b).

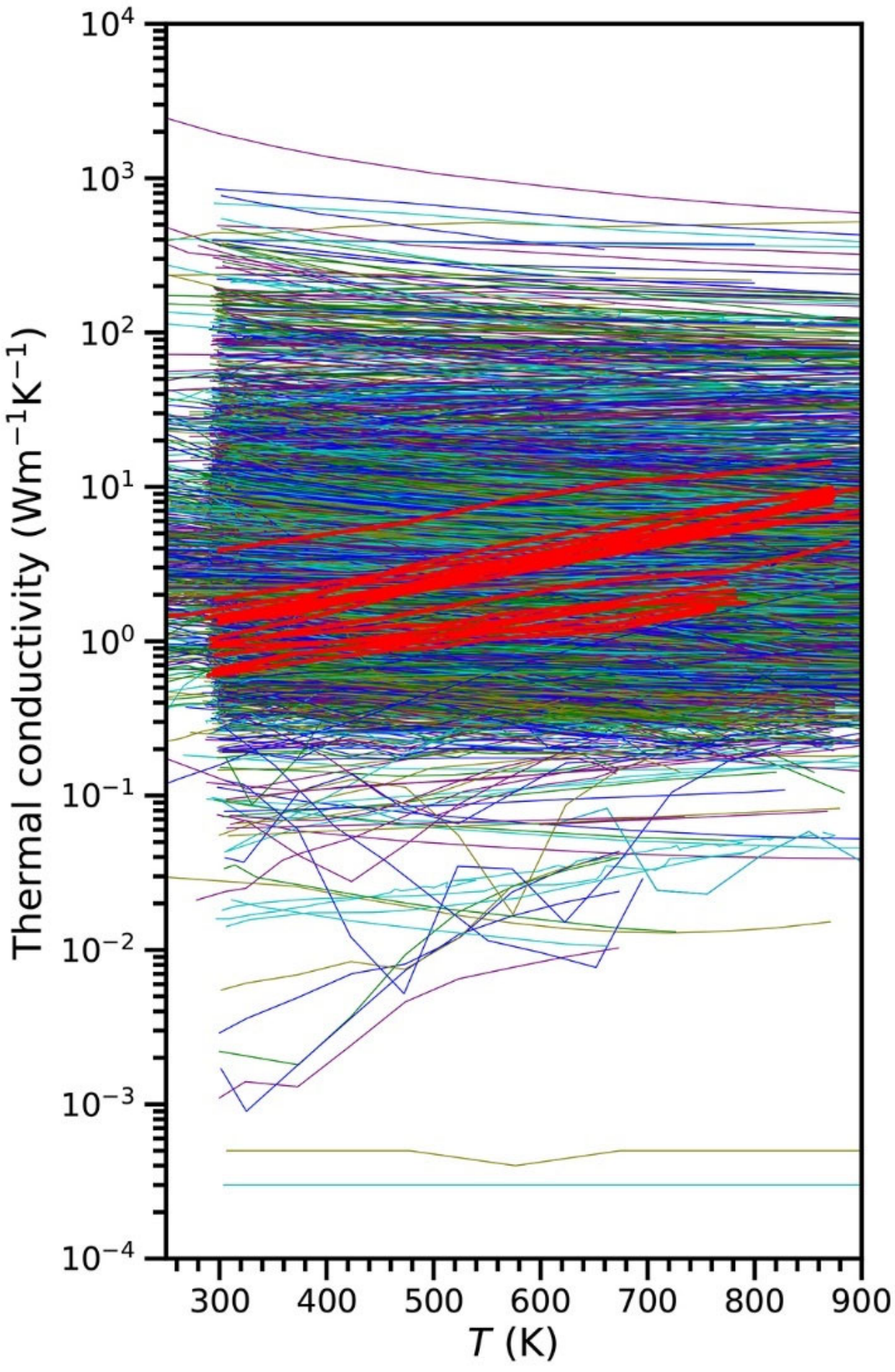

FIG. 3. Temperature dependence of thermal conductivity λ($T$) for all 10,112 data. Red lines indicate λ($T$) for QCs and lines with other colors indicate those for others. Here, except for red, different colors are meaningless; a single color would simply fill a large area because of densely populated lines.

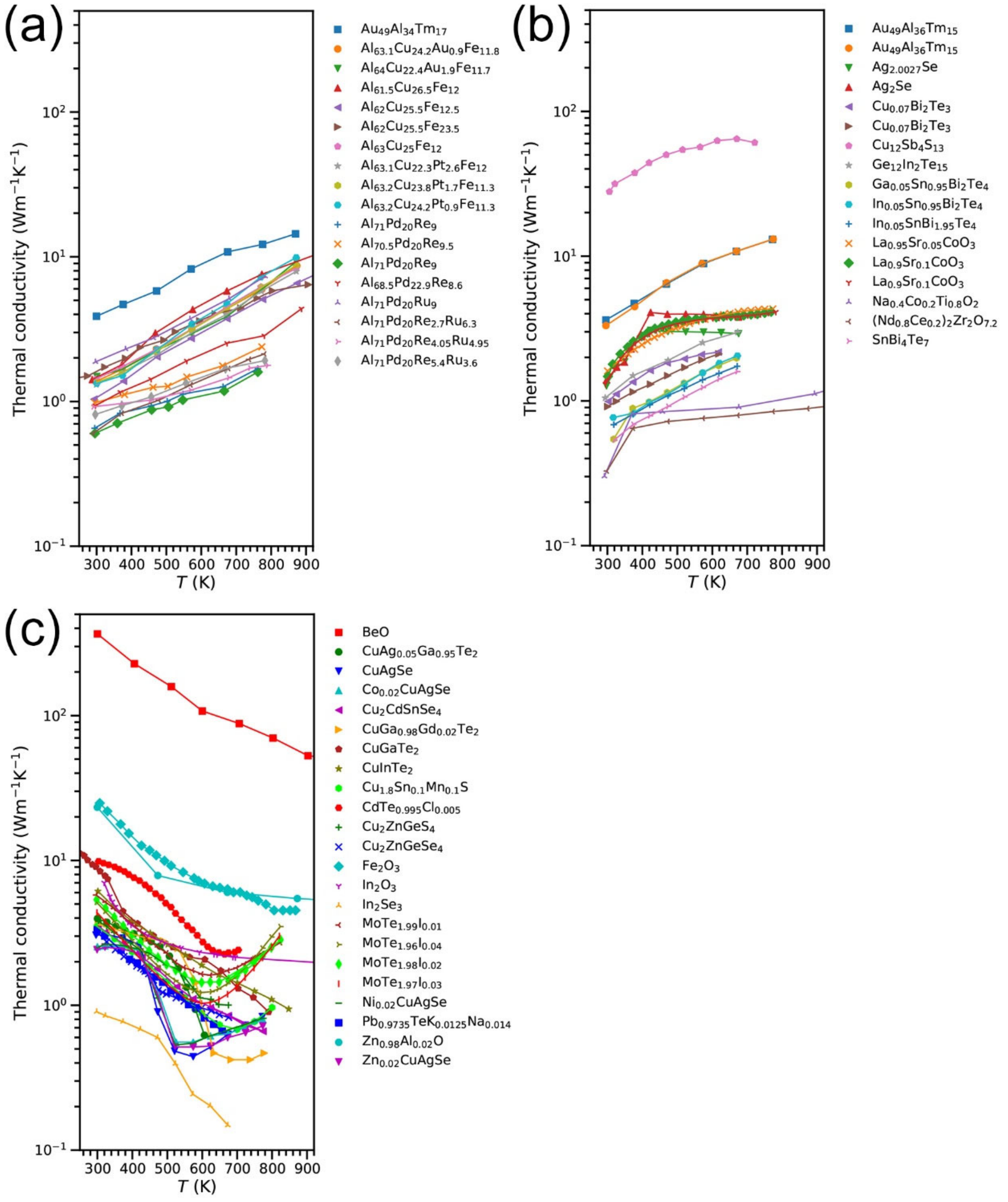

FIG. 4. Temperature dependence of thermal conductivity λ(*T*). (a) λ(*T*) for QCs. (b) λ(*T*) for the samples indicated in black in Table 1. (c) λ(*T*) for the samples in Table 2.

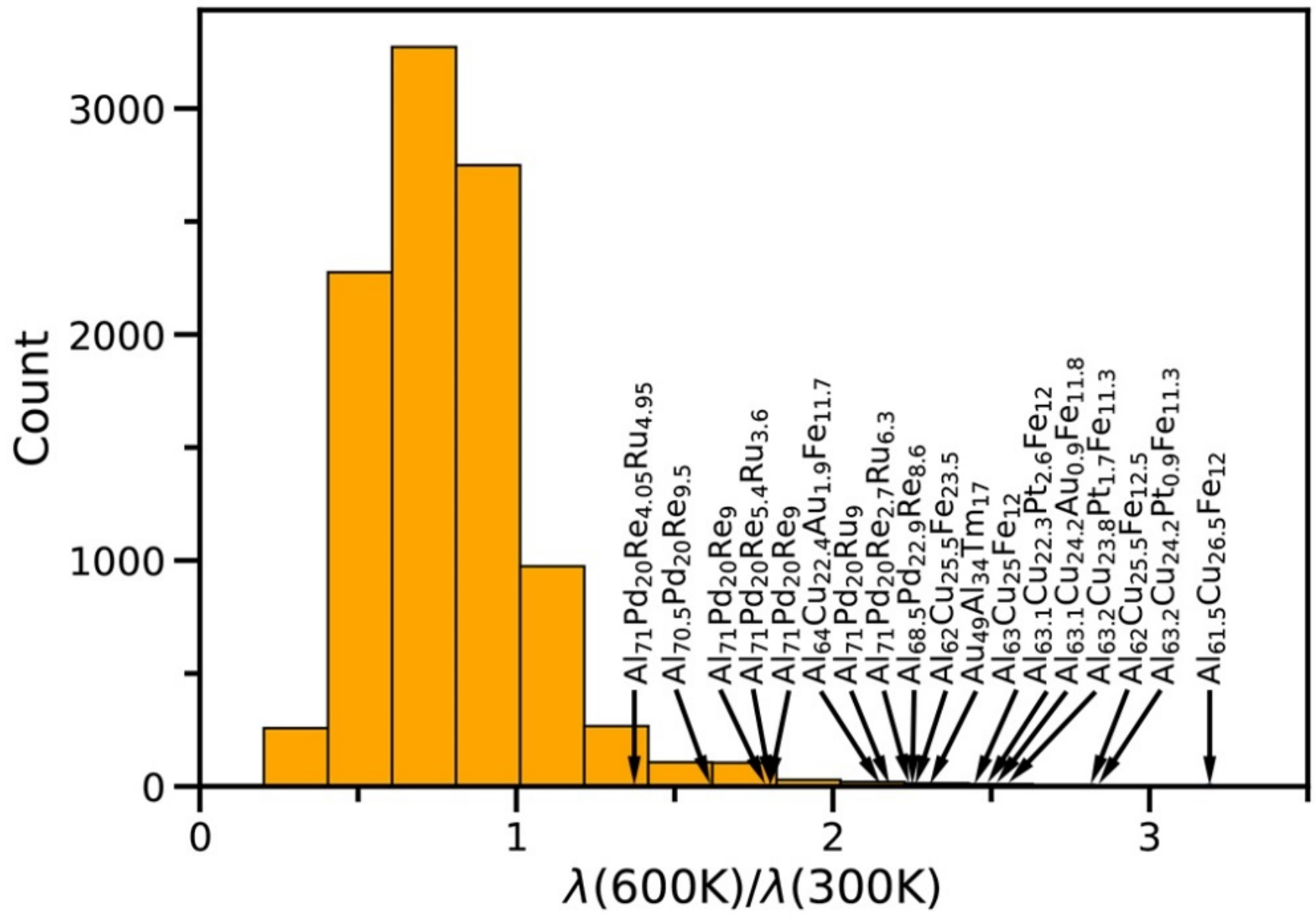

FIG. 5. Distribution of the ratio of the thermal conductivities $R$=λ(600K)/λ(300K). Arrows indicate the $R$ values for QCs.

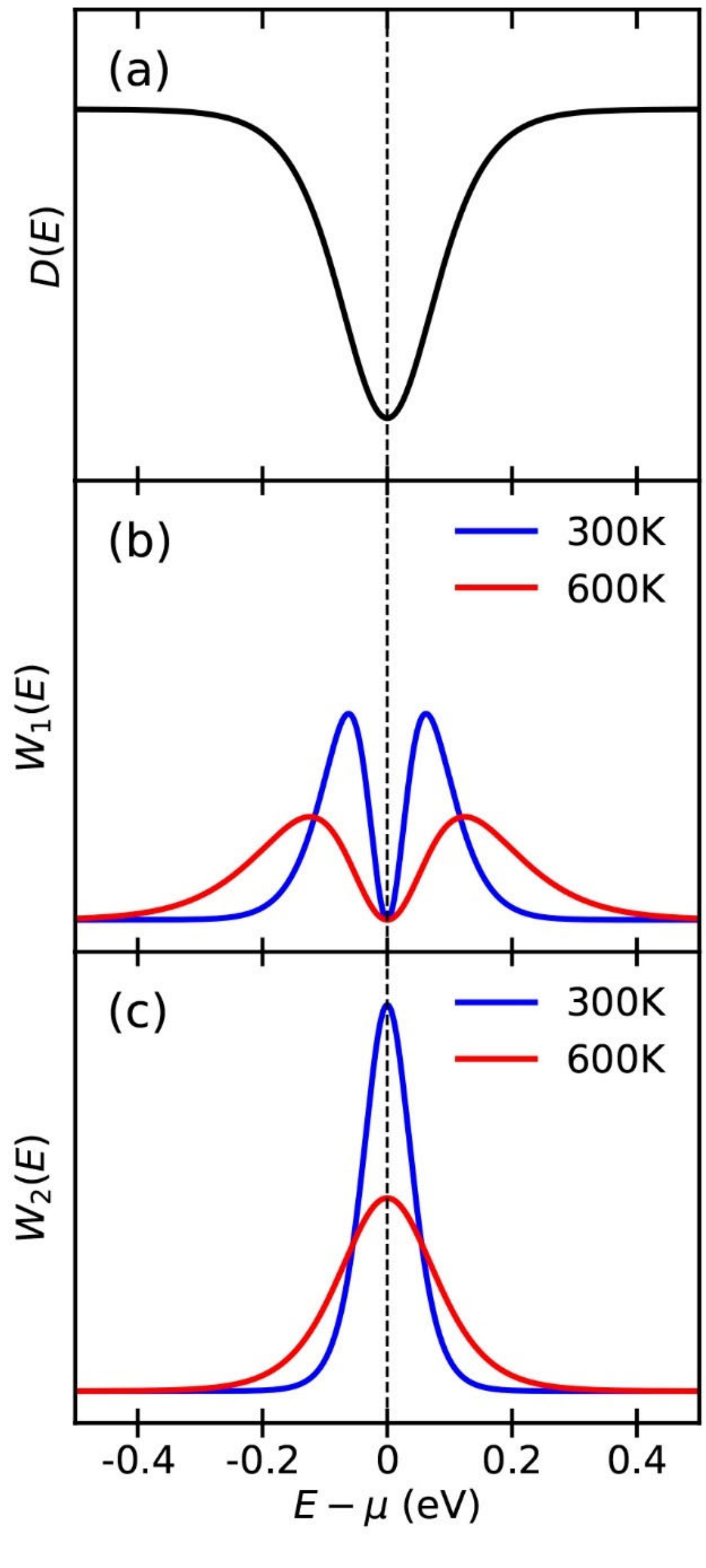

FIG. 6. (a) A schematic illustration of a peudogap in electronic density of states $D(E)$. (b) the window functions $W_1(E, T_0)$ for $T_0$ =300 and 600 K. (c) the window functions $W_2(E, T_0)$ for $T_0$ =300 and 600 K.

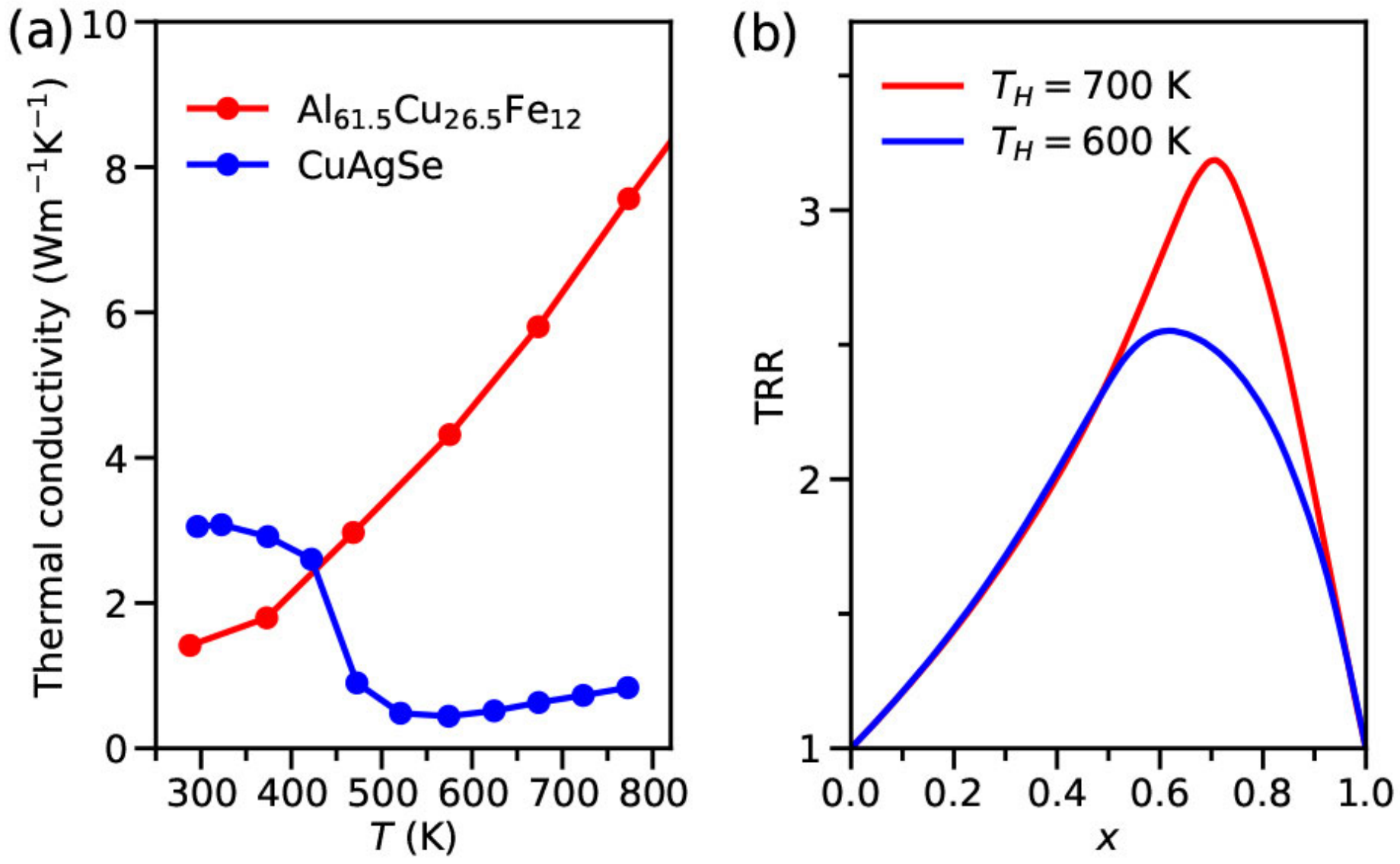

FIG. 7. (a) Temperature dependence of thermal conductivity for QC-$Al_{61.5}Cu_{26.5}Fe_{12}$ and CuAgSe. (b) Calculated thermal rectification ratio (TRR) of the composite thermal diode consisting of $Al_{61.5}Cu_{26.5}Fe_{12}$ as material A and CuAgSe as material B for ($T_H$, $T_L$)=(600 K, 300 K) and ($T_H$, $T_L$)=(700 K, 300 K), plotted against the length ratio $x$.

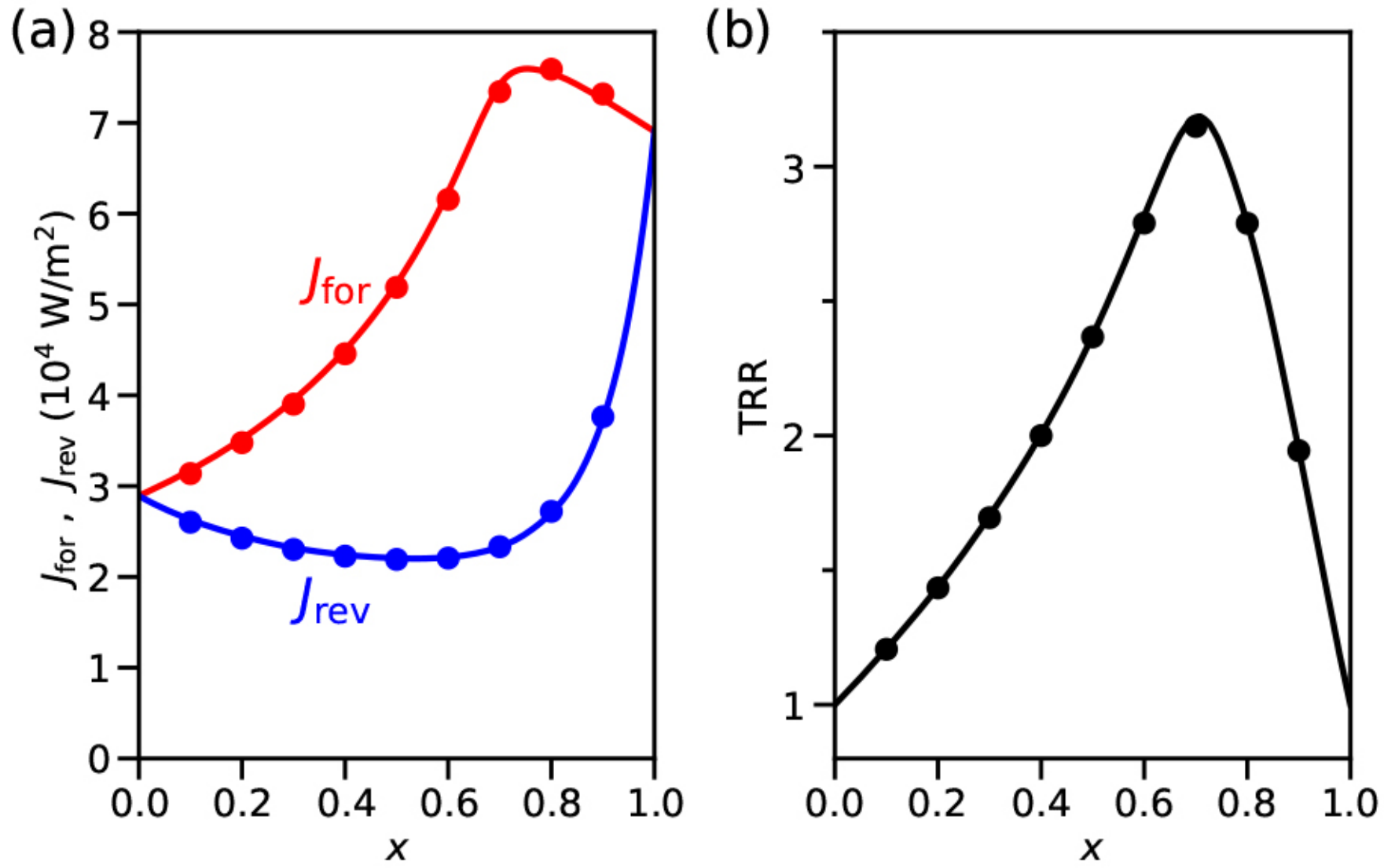

FIG. 8. Comparison of the simulation results (solid circles) with the results (lines) of the analytical calculations by Eqs. (1)-(4) in text for the thermal diode of $Al_{61.5}Cu_{26.5}Fe_{12}$ and CuAgSe with ($T_H$, $T_L$)=(700 K, 300 K). (a) Heat fluxes for the forward and reverse directions as a function of the length ratio $x$. (b) Thermal rectification ratio (TRR) as a function of the length ratio $x$.

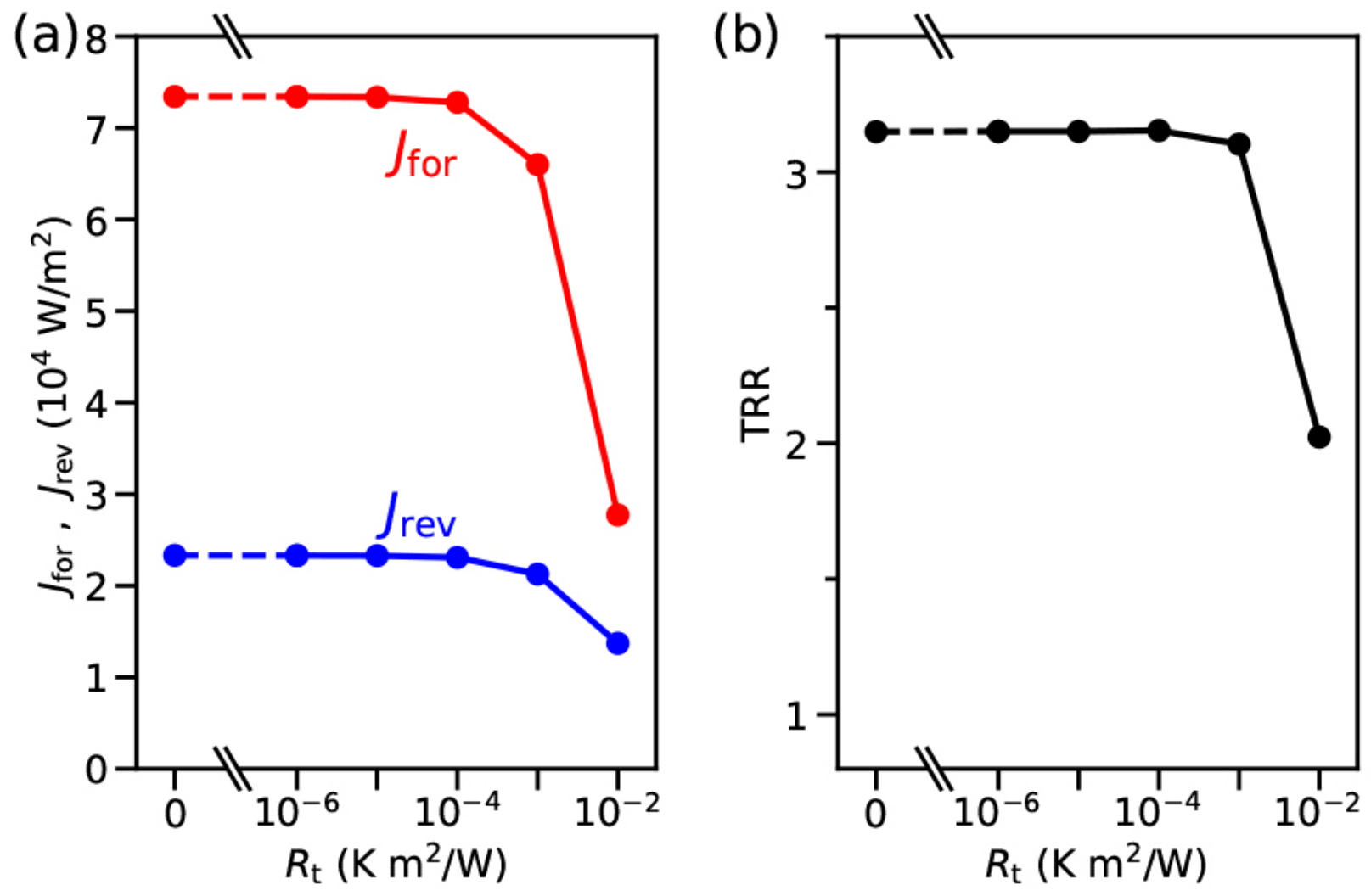

FIG. 9. (a) Heat fluxes for the forward and reverse directions plotted against the thermal resistance $R_t$. (b) Thermal rectification ratio (TRR) plotted against the thermal resistance $R_t$.